\documentclass[12pt,a4paper]{article}
\usepackage{graphicx}
\usepackage{amsmath}
\usepackage{amsfonts}
\usepackage{amssymb}
\usepackage{color}
\usepackage{caption}
\usepackage{units}
\usepackage[bookmarks=true,bookmarksopen=true,bookmarksopenlevel=2]{hyperref} 
\hypersetup{
  pdfpagemode=UseOutlines,
  pdfstartview=Fit,
  pdftitle={},
  pdfsubject={},
  pdfauthor={},
  pdfkeywords={},
  pdfcreator={pdflatex}
}

\begin{document}

\thispagestyle{empty}

\title{An adaptive multiphysics model coupling vertical equilibrium and full multidimensions for multiphase flow in porous media}
\date{November 27, 2017}
\author{Beatrix Becker$^{1}$, Bo Guo$^{2}$, Karl Bandilla$^{3}$, Michael A. Celia$^{3}$,\\ Bernd Flemisch$^{1}$, Rainer Helmig$^{1}$}
\maketitle

\noindent
$^{1}${Department of Hydromechanics and Modelling of Hydrosystems, University of Stuttgart, Stuttgart, Germany}\\
$^{2}${Department of Energy Resources Engineering, Stanford University, Stanford, USA}\\
$^{3}${Department of Civil and Environmental Engineering, Princeton University, Princeton, USA}
%
\section*{Key points}
\begin{itemize}
\item Develop a framework to couple a vertical equilibrium model to a full multidimensional model
\item Develop a local criterion to determine the applicability of vertical equilibrium models
\item Develop an efficient adaptive algorithm to assign subdomains
\end{itemize}
\section*{Abstract}
Efficient multiphysics models that can adapt to the varying complexity of physical processes in space and time
are desirable for modeling fluid migration in the subsurface.
Vertical equilibrium (VE) models are simplified mathematical models that are computationally efficient but rely on the assumption of instant gravity segregation
of the two phases, which may not be valid at all times or at all locations in the domain.
Here, we present a multiphysics model that couples a VE model to a full multidimensional model that has no reduction in dimensionality.
We develop a criterion that determines subdomains where the VE assumption is
valid during simulation. The VE model is then adaptively applied in those subdomains, reducing the number of computational cells
due to the reduction in dimensionality, while the rest of the
domain is solved by the full multidimensional model.
We analyze how the threshold parameter of the criterion influences accuracy and computational cost of the new multiphysics model
and give recommendations for the choice of optimal threshold parameters.
Finally, we use a test case of gas injection to show that the adaptive multiphysics model is much more computationally
efficient than using the full multidimensional model in the entire domain, while maintaining much of the accuracy.
\section{Introduction}
Conventional underground natural gas storage, large-scale CO$_2$ storage, and storage of compressed air or hydrogen in the subsurface
all require numerical models for multiphase flow
to help investigate potential storage sites, determine optimal operational parameters, and ensure safety of operation.
These models often have to deal with large spatial domains in the range of kilometers to hundreds of kilometers in horizontal length
and tens to hundreds of meters in vertical height, and time scales ranging from hours to thousands of years \cite{Nordbotten:2011}.
In addition, due to the uncertainty of geological parameters, a large number of simulations runs (e.g., Monte-Carlo simulation)
may be required for risk assessment.
For this purpose, it is desirable to have computational models that are robust, fast and give an acurate prediction of the system.\par
In practice, there are often multiple physical processes that are of interest and the desired degree of accuracy changes in
space and time.
An example of a system with varying complexity can be found in underground gas storage (e.g., natural gas, compressed air, hydrogen storage
in saline aquifers).
Gas injected into a saline aquifer leads to a two-phase flow system, in which gas moves laterally outward from the injection point
and at the same time upward due to buoyancy.
The overall spatial extent of the gas plume is important in general, but a much more detailed flow field is desired
near the well than farther away, e.g., for well management.
In addition, gas near the well migrates in the vertical as well as horizontal direction during injection and extraction,
while farther away from the well hydrostatic pressure profiles may have developed in the vertical direction.
At these larger distances, the gas may be considered to be in vertical equilibrium (VE) with the brine phase,
which is exploited by so-called VE models that solve vertically integrated equations
and analytically reconstruct the solution in the vertical direction using the VE assumption.
In subdomains where the VE assumption is valid, VE models
give accurate solutions at significantly lower computational costs.
Therefore, a full multidimensional
model may be applied close to the injection well, while a VE model covers the domain
farther from the well.\par
Multiphysics models are robust and computationally efficient on domains with varying complexity because they can adaptively match
model complexity to domain/process complexity for different parts of the domain, which significantly reduces computational costs.
An overview of multiphysics models can be found in \cite{Wheeler20021147} and \cite{HELMIG201352}. Models which couple the transition from one submodel
to another in time are distinguished from models which couple in space.
Coupled models in space can be overlapping within one domain,
e.g., coupling of flow and geomechanics (see \cite{WHITE201655} for a comprehensive framework)
or coupling of different scales. They can also be coupled in separate subdomains with shared interfaces,
e.g., models for flow inside discrete lower-dimensional fractures embedded in the porous matrix
(see, e.g., \cite{singhal2010applied}, \cite{sahimi2011flow} for a comprehensive review)
or compartments with different models coupled in one domain.
Coupling compartments with different models often involves so-called mortar methods \cite{bernardi1989new}, \cite {belgacem1999mortar},
that use Lagrange multipliers at the interfaces to realize the coupling.\par
Another group of coupling schemes exploits similarities between the individual mathematical equations of the subdomains
to couple models without requiring specifically constructed coupling conditions.
One example of such a coupled model was developed in \cite{fritz2012decoupled} for two-phase and single-phase flow.
This multiphysics model considers two-phase multicomponent flow only where both phases are present and solves a simpler one-phase model that is
a degenerate version of the two-phase model everywhere else.
The approach was extended to non-isothermal flow by \cite{Faigle201516}, using a subdomain in which non-isothermal effects are accounted for
and a subdomain where simpler, isothermal equations are solved. The method is shown to be accurate and significantly reduces computational cost.
Another example is presented by \cite{guo2016multiscale}, where multiscale vertically integrated models that can capture vertical two-phase flow dynamics
are coupled for gas migration in a layered geological formation.
The coarse scale consists of several horizontal layers that are vertically integrated.
They are coupled together by formulating a new coarse-scale pressure equation that computes the vertical fluxes between the layers.
In each coarse-scale layer, horizontal and vertical fluxes are determined on the fine scale.
The transport calculation on the fine scale is coupled to the coarse scale sequentially.
While we were finalizing our paper for submission, we learned a recent work on arXiv by \cite{2017arXiv171008735M} who have developed a
multiresolution coupled vertical equilibrium model for fast flexible simulation of CO$_2$ storage. This framework allows the coupling of different dimensions.\par
In this paper we present an adaptive multiphysics framework that allows coupling of spatially non-overlapping compartments
with different models and does not require mortar techniques.
We target gas injection and migration in saline aquifers as an example application.
In a first step we develop a multiphysics model that couples a full  multidimensional two-phase model to a VE model with reduced dimensionality.
The VE model is applied in regions of the domain where the vertical equilibrium assumption holds, while the
full multidimensional model is applied in the rest of the domain.
We design a criterion to adaptively identify the subdomains where the VE model can be applied during the simulation.
For the coupling of the subdomains, we exploit the fact
that all fine-scale variables of the VE model can be reconstructed at every point in the vertical direction. This leads to the introduction of subcells
in the VE grid columns at the interface.
We achieve the coupling through the fluxes across the subdomain boundaries from the full multidimensional cells to the subcells.
The resulting system is solved monolithically.\par
The paper is structured as follows. We first introduce the full multidimensional and the VE model. We then
present our coupled multiphysics model with the calculation of the fluxes across the boundaries between submodels.
Following that, we develop and analyze criteria for vertical equilibrium and present the adaptive algoritm.
Lastly, we show the applicability of our approach on a test case of gas injection in an aquifer and give
recommendations for choosing the optimal threshold parameter for the adaptive algorithm.
\section{Full multidimensional model and VE model}
We first present the general three-dimensional governing equations for two-phase flow in a porous medium.
Then we derive the coarse-scale and fine-scale equations for the VE model by vertically integrating the three-dimensional governing
equations.\par
\subsection{Full multidimensional model: governing equations}
The three-dimensional continuity equation for each fluid phase $\alpha$, assuming incompressible fluid phases and a rigid solid matrix, is:
\begin{equation}
\phi \frac{\partial s_{\alpha}}{\partial t} + \nabla \cdot \mathbf{u}_{\alpha} = q_{\alpha}, \quad \alpha = \mathrm{w, n},
\label{eq:mass}
\end{equation}
where $\phi$ is the porosity [-], $s_{\alpha}$ is the phase saturation [-], $t$ is the time [T], $\mathbf{u}_{\alpha}$ is the Darcy flux [L/T] and $q_{\alpha}$
is the source/sink term [1/T]. The phase $\alpha$ can either be
a wetting phase with a subscript "w" or a nonwetting phase with a subscript "n", resulting in two equations that need to be solved for a two-phase system.
The fluid phases are assumed incompressible for simplicity of presentation.\par
The extension of Darcy's law for multiphase flow states that for each phase $\alpha$,
\begin{equation}
\mathbf{u}_{\alpha} = -\mathbf{k} \lambda_{\alpha} \left(\nabla p_{\alpha} + \varrho_{\alpha} g \nabla z \right), \quad \alpha = \mathrm{w, n},
\label{eq:Darcy}
\end{equation}
where $\mathbf{k}$ is the intrinsic permeability tensor [L$^2$] and $\lambda_{\alpha}$ is the phase mobility [(L T)/M]
with $\lambda_{\alpha} = \frac{k_{\mathrm{r},\alpha}}{\mu_\alpha}$ and $k_{\mathrm{r},\alpha}$ being the relative permeability [-] of phase $\alpha$ which depends on the
wetting phase saturation $s_\mathrm{w}$ and often needs to be determined empirically. $\mu_{\alpha}$ is the viscosity [M/(L T)] of phase $\alpha$,
$p_{\alpha}$ is the phase pressure [M/(L T$^2$)], $\varrho_{\alpha}$ is the phase density [M/L$^3$], $g$ is the gravitational acceleration [L/T$^2$],
and $z$ is the vertical coordinate [L] pointing upward.\par
The two equations (one for each phase) resulting from (\ref{eq:mass}) with Darcy's law (\ref{eq:Darcy}) inserted
are solved for the four primary unknowns, $p_{\alpha}$ and $s_{\alpha}$,
by using the closure equations $s_\mathrm{w} + s_\mathrm{n} = 1$ and $p_\mathrm{c}(s_\mathrm{w}) = p_\mathrm{n} - p_\mathrm{w}$, where $p_\mathrm{c}(s_\mathrm{w})$
is the capillary pressure function, which is assumed to be a function of wetting phase saturation $s_\mathrm{w}$.\par
\subsection{VE model: coarse-scale and fine-scale equations}
A less dense phase injected into a porous formation tends to move upward and segregate from the denser resident phase due to buoyancy.
This buoyant segregation can be used to simplify the governing equations of fluid flow, utilizing the VE assumption
\cite{yortsos1995theoretical}, \cite{lake1989enhanced}.
The VE assumption postulates that the two fluid phases have segregated due to buoyancy, and that the phase pressures have reached gravity-capillary
equilibrium in the vertical direction.
With the VE assumption, the form of the pressure distribution in the vertical direction is known \textit{a priori}.
This can be used to simplify the governing equations of fluid flow by integrating over the vertical direction which leads to a reduction of dimensionality.
The details along the vertical direction can be reconstructed from the imposed equilibrium pressure distribution.
VE models are cast into a multiscale framework by identifying the detailed solution in the vertical direction as the fine scale
and the vertically-integrated equations as the coarse scale \cite{Nordbotten:2011}.
We refer to vertically integrated variables
as 'coarse-scale' variables, denoted by upper-case letters, and to the variations along the vertical direction
as 'fine-scale' variables that are denoted by lower-case letters.\par
In the following we assume that the nonwetting phase is less dense than the wetting phase and that the nonwetting phase consequently
forms a plume below a no-flow upper boundary.
We consider an aquifer with the top and bottom closed to flow.
The full multidimensional governing equation (\ref{eq:mass}) is integrated over the vertical direction from the bottom of the aquifer, $\xi_B$,
to the top of the aquifer, $\xi_T$,
which results in the coarse-scale equations for the VE model:
\begin{equation}
\Phi \frac{\partial S_{\alpha}}{\partial t} + \nabla \cdot \mathbf{U}_{\alpha} = Q_{\alpha}, \quad \alpha = \mathrm{w,\, n},
\label{eq:massCoarse}
\end{equation}
with the depth-integrated parameters:
\begin{eqnarray}
\Phi &=& \int_{\xi_\mathrm{B}}^{\xi_\mathrm{T}} \phi \mathrm dz,\\
S_{\alpha} &=& \frac{1}{\Phi} \int_{\xi_\mathrm{B}}^{\xi_\mathrm{T}} \phi s_{\alpha} \mathrm dz,\\
\mathbf{U}_{\alpha} &=& \int_{\xi_\mathrm{B}}^{\xi_\mathrm{T}} \mathbf{u}_{\alpha,\mbox{\tiny{//}}} \mathrm dz,\\
Q_{\alpha} &=& \int_{\xi_\mathrm{B}}^{\xi_\mathrm{T}} q_{\alpha} \mathrm dz,
\end{eqnarray}
with the subscript '//' denoting the plane of the lower aquifer boundary.
The depth-integrated Darcy flux is found by vertically integrating Darcy's law over the height of the aquifer as:
\begin{equation}
\mathbf{U}_{\alpha} = -\mathbf{K} \Lambda_{\alpha} \left(\nabla P_{\alpha} + \varrho_{\alpha} g \nabla \xi_\mathrm{B} \right), \quad \alpha = \mathrm{w,\, n},
\label{eq:DarcyCoarse}
\end{equation}
with the depth-integrated permeability and depth-averaged mobility:
\begin{eqnarray}
\mathbf{K} &=& \int_{\xi_\mathrm{B}}^{\xi_\mathrm{T}} \mathbf{k}_{\mbox{\tiny{//}}} \mathrm dz,\\
\Lambda_{\alpha} &=& \mathbf{K}^{-1} \int_{\xi_\mathrm{B}}^{\xi_\mathrm{T}} \mathbf{k}_{\mbox{\tiny{//}}} \lambda_{\alpha} \mathrm dz.
\label{eq:coarseRelPerm}
\end{eqnarray}
The coarse-scale pressure $P_{\alpha}$ of phase $\alpha$ in the vertically integrated Darcy's law is defined as the phase pressure at the bottom of the aquifer.\par
Two closure equations are required again to solve for the four unknown primary variables $P_{\alpha}$ and $S_{\alpha}$: $S_\mathrm{w} + S_\mathrm{n} = 1$
and the coarse-scale pseudo
capillary pressure $P_\mathrm{c}(S_\mathrm{w}) = P_\mathrm{n} - P_\mathrm{w}$ that relates the coarse-scale pressure
difference at the bottom of the aquifer to the coarse-scale saturation.
The coarse-scale nonwetting phase pressure at the bottom of the aquifer $P_\mathrm{n}$ is constructed from the linear extension of the pressure distribution for the nonwetting phase
inside the plume to regions below that.\par
After the coarse-scale problem is solved, the fine-scale solution in the vertical direction can be reconstructed based on the
coarse-scale quantities $P_{\alpha}$ and $S_{\alpha}$.
The fine-scale pressure is reconstructed based on the above stated assumption of a hydrostatic pressure profile.
Given the two fine-scale phase pressures at every point in the vertical direction, the fine-scale capillary pressure function $p_\mathrm{c}(s_\mathrm{w})$ can be inverted to
give the fine-scale saturation profile. The fine-scale saturation is used to calculate the fine-scale relative permeability. By integrating the fine-scale relative
permeability using (\ref{eq:coarseRelPerm}), the coarse-scale relative permeability is updated.\par
\section{Coupling VE model with full multidimensional model}
In this section we present the coupling scheme to couple the two models at the interfaces of the subdomains.
The coupling of the subdomains is implemented in a monolithic framework.
We exploit similarities between the full multidimensional governing equations (\ref{eq:mass}, \ref{eq:Darcy}) and the VE
coarse-scale governing equations (\ref{eq:massCoarse}, \ref{eq:DarcyCoarse}), which have the same form.
They balance a storage term consisting of a porosity and
the time derivative of the saturation with fluxes and a source/sink term, while
the flux term is calculated with the gradient of driving forces, pressure and gravity, multiplied by a term describing flow resistance.
In the case of the VE model the quantities are depth-integrated over the height of the VE column.
In the following we present the formulation of fluxes across the subdomain interfaces and the computational algorithm for the
coupled multiphysics model.\par
\subsection{Fluxes across subdomain boundaries}
\begin{figure}[!htb]
\begin{center}
\includegraphics[width=0.28\textwidth]{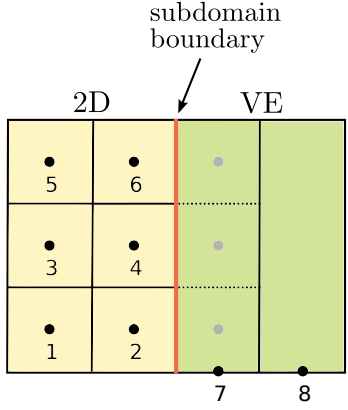}
\caption{Schematic of the computational grid with subcells (dotted lines) at the interface between two subdomains.
Black dots denote the calculation points of the primary variables in both subdomains.
Gray dots denote the calculation points of the primary variables for subcells, which can be seen as fine-scale cells of the VE model.}
\label{grid}
\end{center}
\end{figure}

We discretize space with a cell-centered finite-volume method.
Figure \ref{grid} shows a possible configuration of grid cells with two subdomains and one shared boundary between them.
In this example we consider a two-dimensional domain, where the first two grid columns are part of the full multidimensional
subdomain, the third and fourth grid column are part of the VE subdomain.
The black dots indicate the location of the calculation points for the primary
variables. For the full multidimensional model the calculation points are located in the cell center,
for the VE model at the bottom of the domain.\par
Fluxes between two full multidimensional cells and between two VE cells are determined using a two-point flux approximation.
For the calculation of fluxes across the subdomain boundary,
the VE grid column directly adjacent to the subdomain boundary is refined into full multidimensional subcells
in the vertical direction, with each subcell corresponding to a neighboring full multidimensional cell (see gray dots and dotted lines in Figure \ref{grid}).
Fluxes are formulated across the interface between each full multidimensional cell and the neighboring VE subcell.
For each full multidimensional cell there is only one flux across the interface to the adjacent VE subcell.
The flux over the subdomain boundary to the VE cell is computed as the sum of the fluxes from the neighboring full
multidimensional cells.\par
In the following we exploit the fact that the primary variables at the calculation points of the VE subcells can be expressed
analytically via the primary variable of the VE cell.
This is because the solution in the vertical direction can be reconstructed analytically using the VE assumption in the VE subdomain.
With this, the total normal velocity from a full multidimensional cell denoted with superscript 'i' to a VE subcell
denoted with superscript 'j*' can be constructed as:
\begin{equation}
v_\mathrm{tot}^{ij*} = v_\mathrm{w}^{ij*} + v_\mathrm{n}^{ij*} = -k^{ij*} \lambda_\mathrm{tot}^{ij*} \left(\frac{p_\mathrm{w}^{j*} - p_\mathrm{w}^{i}}{\Delta x}
+ f_\mathrm{n}^{ij*} \frac{p_\mathrm{c}^{j*}-p_\mathrm{c}^{i}}{\Delta x}
             + f_\mathrm{w}^{ij*} \varrho_\mathrm{w} g \nabla z + f_\mathrm{n}^{ij*} \varrho_\mathrm{n} g \nabla z \right),
\end{equation}
where $p_\mathrm{w}^{j*}$ is the reconstructed pressure
and $p_\mathrm{c}^{j*}$ the reconstructed capillary pressure at the calculation point of the VE subcell.
The reconstructed $p_\mathrm{w}^{j*}$ and $p_\mathrm{c}^{j*}$ have the following form:
\begin{equation}
\begin{split}
p_\mathrm{w}^{j*}&=P_\mathrm{w}^j - \varrho_\mathrm{w} g z^*,\\
p_\mathrm{c}^{j*}&=p_\mathrm{c}(s_\mathrm{w}^{j*}),
\end{split}
\end{equation}
with $P_\mathrm{w}^j$ the coarse-scale pressure of the VE cell, $z^*$ the z-coordinate of the calculation point of the VE subcell,
and $s_\mathrm{w}^{j*}$ the reconstructed wetting phase saturation at the node in the middle of the subcell.
The mobilites for the VE subcells are based on
the average wetting phase saturation within the subcell, $\bar{s}_\mathrm{w}^*$.
We apply full upwinding for the mobilities, so that they are either taken directly from the full multidimensional cell or the reconstructed subcell in the VE column:
\begin{equation}
\begin{split}
\lambda_{\alpha}^{ij*}=\begin{cases}
  \lambda_{\alpha}^{i}(s_\mathrm{w}^{i}) &\text{if} \quad p_{\alpha}^{i} > p_{\alpha}^{j*},\\
  \lambda_{\alpha}^{j*}(\bar{s}_\mathrm{w}^{j*}) &\text{if} \quad p_{\alpha}^{i} < p_{\alpha}^{j*}.\\
\end{cases}
\end{split}
\end{equation}
For $p_{\alpha}^{i} = p_{\alpha}^{j*}$, the flux is zero and the choice of mobility does not matter.\par
This concept is used for all cells at the interface between the full multidimensional subdomain and the VE subdomain.
The flux from a VE cell to neighboring full multidimensional cells is determined as the sum of the individual fluxes
over the subdomain boundary from VE subcells to full multidimensional cells.
Following this approach, all fluxes are the same as calculated from either the VE cell side or the full multidimensional cell sides.\par
\subsection{Computational algorithm}
The full multidimensional governing equations (\ref{eq:mass}, \ref{eq:Darcy}) and the VE
coarse-scale governing equations (\ref{eq:massCoarse}, \ref{eq:DarcyCoarse}),
can in principle be solved fully implicitly, sequentially implicitly or sequentially with a combination of implicit and explicit schemes.
Here, we reformulate the governing equations into a pressure and a saturation equation and solve them sequentially
with an implicit pressure, explicit saturation (IMPES) algorithm.
The pressure equation is solved in an implicit manner with a single computational matrix for the entire domain
(multidimensional plus VE subdomains) and therefore
we do not use iterations between subdomains.
Specifically this means that the pressure in the different subdomains is solved simultaneously and
each VE cell and each full multidimensional cell contribute one row to the pressure matrix.
For cells at the subdomain interfaces the velocity is constructed as shown above with the help of the VE subcells.
For VE cells at the subdomain
boundary the fluxes from all neighboring cells are taken into account.
Once the fluxes have been calculated from the pressure solution, the saturations are updated explicitly for each cell
using the saturation equation.
The corresponding saturation equation for VE cells at the subdomain interface again takes into account
all fluxes from neighboring full dimensional cells and VE cells.\par
The saturation is lagged one time step in the IMPES algorithm,
meaning the same capillary pressure as well as all other secondary variables that depend on the saturation
are used in the pressure and saturation step of the algorithm.
The phase fluxes are computed from the pressure field solved in the pressure step, resulting in a mass-conservative scheme.
We note that the IMPES algorithm assumes a weak coupling between pressure and saturation equations.
If saturations change significantly during one time step, iterating between
the pressure and saturation step of the IMPES algorithm or a fully implicit scheme becomes necessary.
This applies regardless of the coupling within the domain.\par
%
\section{Criteria for vertical equilibrium and adaptive coupling}
In this section we first discuss general criteria for vertical equilibrium. We then develop criteria to determine
when and where to apply a VE subdomain in the multiphysics model and analyze their behavior.
In a third step, we present an algorithm that adaptively moves the boundaries between subdomains.\par
\subsection{Criteria for vertical equilibrium}
We identify two groups of criteria that determine whether the vertical equilibrium assumption holds:
one is referred to as a global criterion and the other is referred to as a local criterion.
The global criterion gives an \textit{a priori} estimate of the time after which the vertical equilibrium assumption
holds in the entire spatial domain, e.g., the segregation time $t_\mathrm{seg}$ \cite{nordbotten2011impact}:
\begin{equation}
t_\mathrm{seg} = \frac{H \phi (1-s_\mathrm{wr}) \mu_\mathrm{w}}{k_\mathrm{r,w} k_\mathrm{v} g (\varrho_\mathrm{w} - \varrho_\mathrm{n})},
\label{eq:tseg}
\end{equation}
with $H$ the height of the aquifer, $s_\mathrm{wr}$ the residual wetting phase saturation and $k_\mathrm{v}$ the vertical component
of the permeability tensor. Practically, we chose the characteristic value for the wetting phase relative permeability $k_\mathrm{r,w}$ to be 1, which leads
to a smallest possible segregation time.
The VE model gives accurate results for time scales that are much larger than the segregation time.\par 
The local criterion can determine if the VE assumption is valid for a specific point in time and space.
This is usually an \textit{a posteriori} criterion, which means the information
is only known during runtime based on the computed solution,
unlike the global criterion which can be evaluated before the solution is computed. As a result the global criterion is
typically very approximative in nature.
Additionally, in a realistic geological setup, there may be regions of the model domain (e.g., near the well, local heterogeneities)
where the VE assumption does not hold, even provided the simulation time is much larger than the segregation time.
Around the well the fluid phases will not reach vertical equilibrium at any time, especially considering frequently alternating injection and extraction cycles.
Because the global criterion gives average information for the entire domain,
it is unsuited to identify local regions where the vertical equilibrium assumption holds.
We will therefore use the local criterion to determine the applicability of the VE model for each vertical grid column in the domain at
every time step.\par
We develop two local criteria that can be used to determine if the wetting and nonwetting phases in a grid column have reached vertical equilibrium.
For the first criterion we compare the full multidimensional profile of wetting-phase saturation in the vertical direction to the VE-profile
that would develop if the phases were segregated and hydrostatic pressure conditions had been reached.
The difference indicates how far away the two fluids are from vertical equilibrium.
The approach for the second criterion is the same, except that we compare relative-permeability profiles of the wetting phase.
For non-linear relative permeability functions the relative-permeability profile will differ from the saturation profile and the two local criteria will
give different values.
Although the saturation profile is directly linked to the state of vertical equilibrium in the column,
the relative permeability profile is more relevant to the calculation of the coarse-scale relative permeability and thus applies more directly to the
development of the plume.\par
The VE-profiles are reconstructed from the total volume of gas inside the grid column, in the same way as the fine-scale
saturation and relative permeability profiles in the VE model are constructed.
\begin{figure}[!htb]
\centering
        \includegraphics[width=0.65\textwidth]{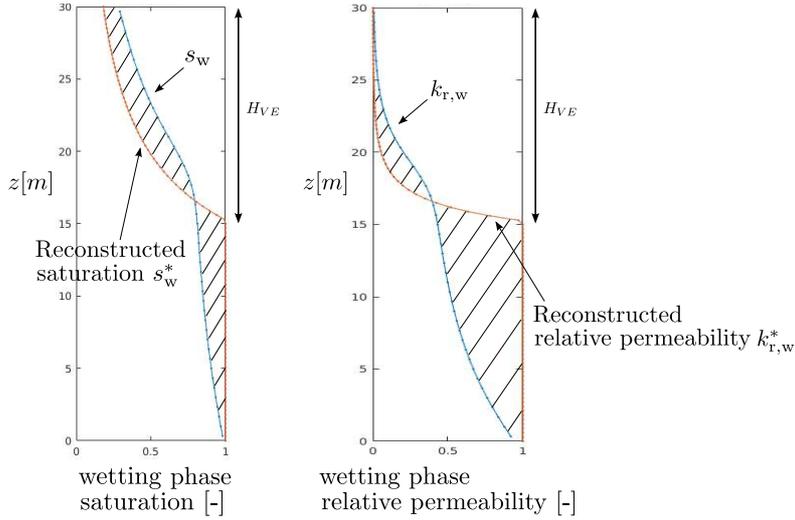}
        \caption{Vertical profiles in one column. Left: wetting phase saturation, right: wetting phase relative permeability
        using a Brooks-Corey relationship with pore size distribution index $\lambda = 2.0$ and entry pressure $p_e = 10^5$~Pa.
        The blue curve is the result of the full multidimensional solution, and
        the orange curve is the reconstructed profile that would develop if the fluid phases were in equilibrium.
        The difference between profiles is depicted as striped areas.}
        \label{criterionProfile}
\end{figure}
We can compute the area of the differences between the profiles (see figure \ref{criterionProfile}) and use that to develop the criteria.
Specifically, we normalize the computed area with the height of the VE-profile
and define $c_\mathrm{sat}$ and $c_\mathrm{relPerm}$ as the criteria values for saturation and relative permeability respectively:
\begin{eqnarray}
c_\mathrm{sat} &=& \frac{\int_{\xi_\mathrm{B}}^{\xi_\mathrm{T}} |s_\mathrm{w} - s_\mathrm{w}^*|\mathrm{dz}}{H_\mathrm{VE}},\\
c_\mathrm{relPerm} &=& \frac{\int_{\xi_\mathrm{B}}^{\xi_\mathrm{T}} |k_\mathrm{r,w} - k_\mathrm{r,w}^*|\mathrm{dz}}{H_\mathrm{VE}}.
\label{eq:crit}
\end{eqnarray}
The vertical equilibrium assumption can be considered to be valid in a grid column during a time step
when the criterion value $c_\mathrm{crit}$ is smaller than a threshold value $\epsilon_\mathrm{crit}$, where the threshold
is a constant value that has to be chosen by the user.\par
\subsection{Criteria analysis}
We analyze the behavior of the two local criteria for vertical equilibrium over space and time as well as for different simulation parameters.
In our two-dimensional test case we inject methane (CH$_4$) from the left over the entire thickness (30~m) of an initially brine-saturated domain.
Bottom and top are closed to flow and Dirichlet conditions are prescribed on the right-hand side with $s_\mathrm{w} = 1.0$ and
a hydrostatic distribution of the brine phase pressure $p_\mathrm{w}$, starting with $10^{7}~\mathrm{Pa}$ at the top.
We chose our domain long enough so that the gas will not reach the right-hand side boundary during the simulation.
We assume a density of $59.2~\mathrm{kg/m}^3$ and a viscosity of $1.202 \times 10^{-5}~\mathrm{Pa}\,\mathrm{s}$ for CH$_4$.
The density of the brine phase is assumed to be $991~\mathrm{kg/m}^3$ and the viscosity of the brine phase
$5.23 \times 10^{-4}~\mathrm{Pa}\,\mathrm{s}$.
We uniformly inject $0.0175~\mathrm{kg/s/m}$ CH$_4$ for 240~h.
The permeability is assumed to be $2000~\mathrm{mD}$ and the porosity $0.2$. For relative permeability and capillary pressure we use Brooks-Corey
curves with pore size distribution index $\lambda = 2$ and entry pressure $p_\mathrm{e} = 10^{5}~\mathrm{Pa}$.
We use a grid resolution of 1~m in both horizontal and vertical direction,
and estimate the segregation time with (\ref{eq:tseg}) as $t_\mathrm{seg} = 48$~h.\par
\begin{figure}[!htb]
\centering
\includegraphics[width=1.0\textwidth]{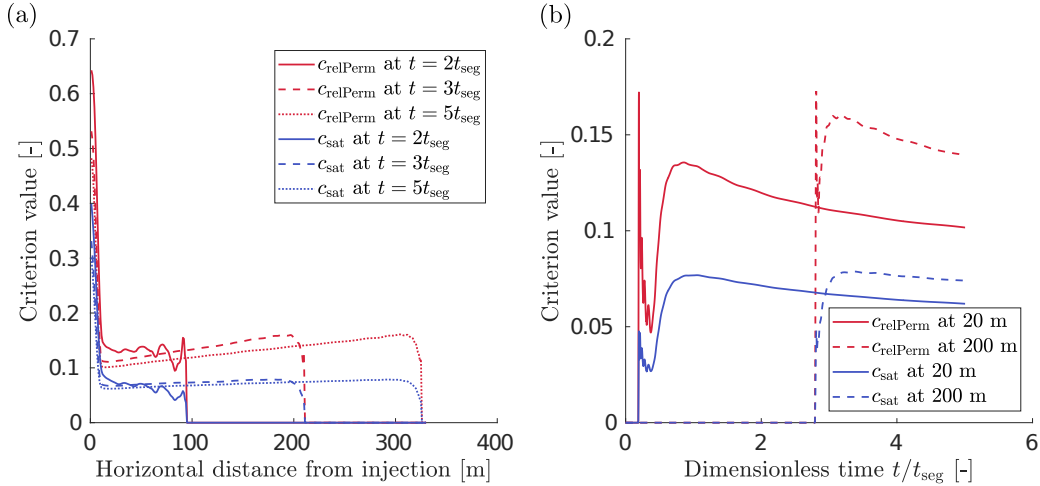}
\caption{Criterion values for the criteria based on saturation and relative permeability respectively. (a)
over the entire length of domain for three times: $2t_\mathrm{seg}$, $3t_\mathrm{seg}$ and $5t_\mathrm{seg}$,
(b) over dimensionless time for a grid column with $20~\mathrm{m}$ distance from the injection location
and a grid column with $200~\mathrm{m}$ distance from the injection location.}
\label{plotCrit}
\end{figure}

We compute the criterion values in each numerical grid column of the two-dimensional domain and compare them in space and time,
as shown in Figure \ref{plotCrit}. We plot criterion values of both criteria over the length of the two-dimensional domain for
three different times and over dimensionless time $t/t_\mathrm{seg}$ for two vertical grid columns.
For our chosen injection scenarios the criterion based on relative permeability gives higher values than the saturation criterion.
This is due to the strong nonlinearity of the relative permeability function.
On one hand, brine held back inside the region of the gas plume
contributes less to the relative permeability criterion than to the saturation criterion since the relative permeability will
be almost zero for small wetting phase saturations.
On the other hand, small gas phase saturations below the gas plume in equilibrium lead to very large criterion values of the relative permeability criterion.
This makes the relative permeability profile deviate strongly from the VE profile in the region below the VE gas plume.
In conclusion, if the gas plume in equilibrium is small compared to the height of the aquifer, the relative permeability criterion will lead
to higher values than the saturation criterion.\par
Around the injection location both criteria show very high values (Figure \ref{plotCrit} (a)). Here the brine and gas phase are not in equilibrium during the simulation
since the gas phase moves continuously upward during injection.
Farther away from the injection point the criteria values decrease steeply which shows that the two phases are much closer to
equilibrium. The saturation criterion stays constant over most of the length of the plume while the
relative permeability criterion increases slightly toward the leading edge of the plume. This is due to the decreasing thickness of the
plume toward the leading edge which, as explained, is penalized more by the relative permeability criterion.\par
For early simulation times a non-monotonic behavior of both criteria can be observed over the length of the domain in Figure \ref{plotCrit} (a).
This is due to the small thickness of the plume in early times and the grid discretization. In some parts of the domain
the vertical location of the gas plume will correspond well with the vertical spacing of the computational grid,
while in others the saturation will be more smeared out due to the finite size of the grid cells. This effect grows less important as the gas plume height
increases with time and contains an increasing number of cells in the vertical direction.
Normalizing both criteria with the height of the VE plume leads to a large peak in criterion values when the leading edge of the gas plume moves
into a cell that has previously been fully saturated with brine.
This can be observed in Figure \ref{plotCrit} (b) for early times. The peak in criterion value is followed by a non-monotonic decrease for both criteria.
This is again due to the finite size of the cells and a simultaneous increase of VE gas plume height.\par
We define requirements for good \textit{a posteriori} criteria for vertical equilibrium, to compare the two local criteria that we developed.
In practice, a good local criterion for vertical equilibrium should:
\begin{itemize}
\item locally start at a high value for early injection times and decrease over time steeply until $t_{sim}>t_{seg}$, then tend toward zero;
\item show enough difference in value when comparing grid columns close to injection and far away from it.
\end{itemize}
Both criteria fullfill the second requirement with very large differences in criterion values at the injection and farther away from it
(Figure \ref{plotCrit} (a)).
For a fixed location in space as in Figure \ref{plotCrit} (b) the criteria values seem to flatten out with time and it appears that they converge to a low,
non-zero value. This value is defined by the finite grid size in the vertical direction beyond which the approximation of the vertical profile
cannot be further improved.
In comparison, the criterion based on the relative permeability shows an overall more promising behavior. It decreases faster for earlier times
and it shows differences also when comparing values at the leading edge and the middle of the plume.\par
Since the local criterion depends on the simulation result of the full multidimensional model, the results depend on the resolution of the grid.
If the grid is too coarse,
it will take longer for the brine phase to drain out of the plume because part of the gas will be smeared out over the grid cells by numerical diffusion.
This inaccuracy will directly be reflected by the local criterion because the profiles in the vertical direction will not resemble vertical equilibrium.
In those cases, higher criterion values can occur although in a real scenario the two phases may already be in equilibrium.
The local criterion is only able to give information about the real physical behavior of the system
when grid resolution is fine enough and the full multidimensional model gives accurate enough results.\par
%
%
%
%
\subsection{Adaptive coupling}
During the simulation, regions where the VE assumption is valid can appear or disappear and change in location and size.
At the beginning of injection, the less dense nonwetting phase is usually not in equilibrium with the denser wetting phase.
Over time, the wetting phase drains out of the plume and the area where the VE model can be applied increases. Around the well, the flow field
will always have components in the vertical direction and require a full multidimensional resolution at all times.
Furthermore, even an already segregated plume can reach heterogeneous
zones that require a full multidimensional resolution for accuracy. An efficient model therefore adapts automatically to changes during the simulation.\par
\begin{figure}[!htb]
\centering
\includegraphics[width=0.27\textwidth]{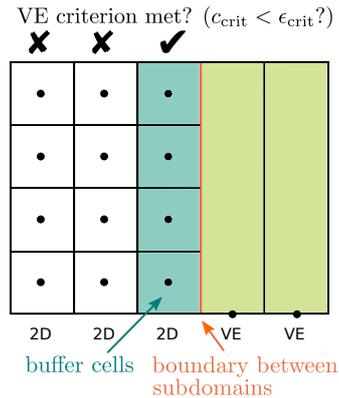}
\caption{Buffer zone between full multidimensional subdomain and VE subdomain: one (or several) grid column(s) that fullfill
the requirement of vertical equilibrium (according to the applied criterion) but are still kept as full multidimensional grid columns to detect changes
in the flow field.}
\label{buffer}
\end{figure}

We develop an algorithm for adaptation that applies the VE model in all regions where the VE assumption is valid and tests the validity in every time step.
The location of the boundaries between the two submodels are found based on the local criteria for vertical equilibrium from the last time step.
The local criterion is evaluated for each full multidimensional grid column before each time step.
If the criterion value $c_\mathrm{crit}$ is smaller than a user-defined threshold value $\epsilon_\mathrm{crit}$,
the grid column is assumed to be in vertical equilibrium.
Depending on the criterion value and the threshold, either one of the following decisions is made for each grid column:
\begin{itemize}
\item A full multidimensional grid column stays full multidimensional if the VE criterion is not met ($c_\mathrm{crit} \geq \epsilon_\mathrm{crit}$).
\item A full multidimensional grid column is turned into a VE column if the VE criterion is met ($c_\mathrm{crit} < \epsilon_\mathrm{crit}$)
and the column is not a direct neighbor to a column where the criterion is not met.
\item A VE grid column is turned into a full multidimensional grid column if it is a direct neighbor to a column where the criterion is not met.
\end{itemize}
The criterion value is used directly to turn full multidimensional grid columns into VE columns. The third requirement from
above is required to turn VE columns back into full multidimensional cells.
Together with the second requirement it results in a buffer zone around the VE columns,
which is made up of full multidimensional columns (see figure \ref{buffer}).
It guarantees that a VE column is converted back to a full multidimensional column before the flow field returns to full multidimensional
at this location.
This approach with one layer of buffer cells assumes that the subdomain boundary does not
need to be moved more than one cell into the horizontal direction in each time step, which should be guaranteed by
fulfilling the CFL criterion of the explicit time stepping.
For stability reasons the buffer zone could be extended to have more than one layer.\\
\section{Results and discussion}
In this section, we use a heterogeneous test case to test accuracy, robustness and computational efficiency of the multiphysics model.
We compare the solution against results from a full multidimensional model and a VE model.
Based on the comparison, we develop guidelines for the choice of the threshold value $\epsilon_{crit}$.
The multiphysics model and the full multidimensional model are
both implemented in DuMu$^{\mathrm{x}}$ \cite{Flemisch20111102}, \cite{ackermann_sina_2017_439488}.\par
\begin{figure}[!htb]
\centering
\includegraphics[width=0.6\textwidth]{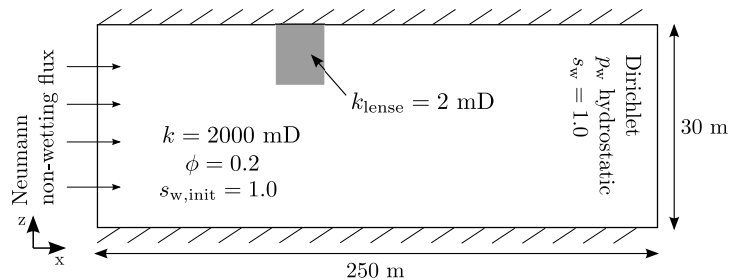}
\caption{A test of gas injection for the adaptive model.
A low-permeability lense is located at the top of the aquifer at $100~\mathrm{m}$ distance from the injection location.}
\label{setup}
\end{figure}

In the test case, we again inject CH$_4$ into a previously brine-saturated domain (see Figure \ref{setup}).
We uniformly inject $0.0175~\mathrm{kg/s/m}$ gas for 192~h.
The scenario is equal to the one used to analyze the VE criterion values with the same geological parameters and the
same fluid properties.
Additionally a low-permeability lense directly below the top boundary of the aquifer is added ($k_\mathrm{lense} = 2~\mathrm{mD}$).
The lense is located at $100~\mathrm{m}$ distance from the injection location with a length of $20~\mathrm{m}$ and a height of $10~\mathrm{m}$.
The entry pressure of the low-permeability lense is kept the same as inside the domain. The gas will pool in front of the lense and
flow around it while a small part of the gas may migrate into the lense. This
creates a full multidimensional flow pattern that can only be resolved accurately with a full multidimensional simulation.
For simplicity, we solve the injection scenarios in two dimensions (horizontal and vertical directions).
However, we note that this is not a necessity for the coupling algorithm.
We choose a grid resolution of 1~m in the horizontal direction and
(in the full multidimensional subdomain) 0.23~m in the vertical direction and apply the relative permeability criterion
to identify subregions.
We vary the threshold value between $0.01$ and $0.06$ to analyze its influence on the simulations
and develop recommendations for the choice of the threshold value.
A full multidimensional solution is obtained on a
two-dimensional grid with a grid resolution of 1~m in the horizontal direction and 0.23~m in the vertical direction.\par
\subsection{Comparison between models}
\begin{figure}[!htb]
\centering
\includegraphics[width=0.5\textwidth]{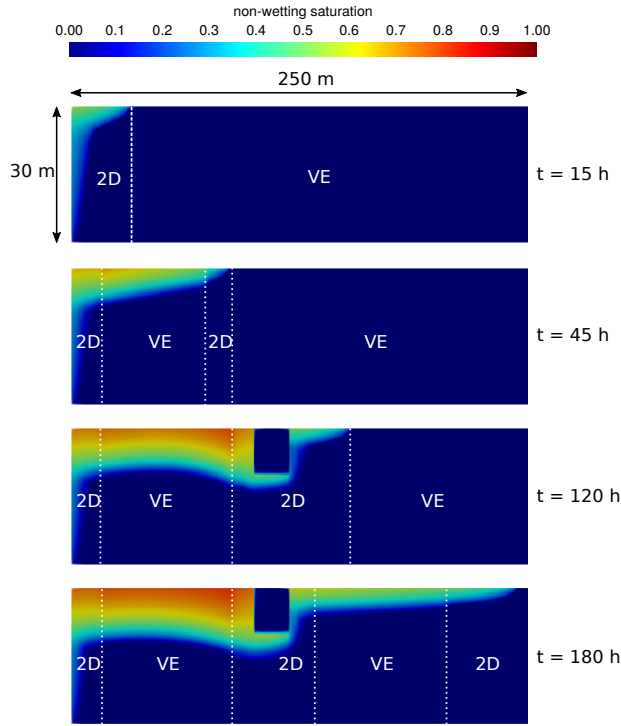}
\caption{Gas phase distribution for the adaptive multiphysics model with a threshold value of $0.03$ for a series of simulation times after injection.
Subdomain boundaries are marked by dotted lines. The reconstructed solution is shown in the VE subdomains .
The domain is homogeneous except for a low-permeability lense at $100~\mathrm{m}$ distance from the injection location.}
\label{satRecons}
\end{figure}

We show the resulting gas phase saturation distribution of the adaptive multiphysics model with a threshold value of 0.03 for different times in Figure \ref{satRecons}.
At the beginning of simulation, only the single-phase region is turned into a VE subdomain by the adaptive algorithm,
which means that the entire gas plume is located within the full multidimensional subdomain.
After a few simulated hours, a second VE subdomain starts developing in the middle of the plume where, according to the criterion, the two fluid phases
have reached vertical equilibrium. When the plume reaches the low-permeability lense, a full multidimensional
region develops around it and accurately captures the flow of gas around the obstacle.
Farther away from the lense, another VE subdomain develops after some time.
During the entire simulation, the area around the
injection stays a full multidimensional subdomain as expected.
The advancing thin leading edge is always resolved in full dimensions as well, because water constantly drains out of it.
The full multidimensional region around the leading edge of the plume
serves here as an indicator for heterogeneous regions like the lense, that would otherwise not be recognized.\par
\begin{figure}[!htb]
\centering
\includegraphics[width=0.5\textwidth]{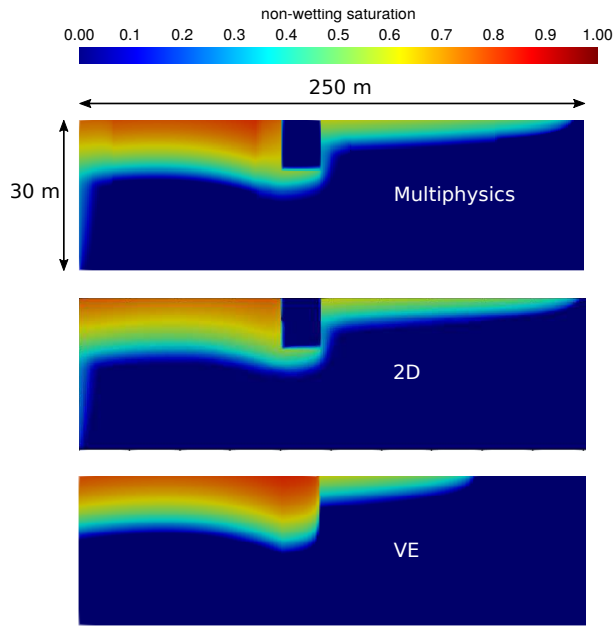}
\caption{Gas phase distribution for the adaptive multiphysics model with a threshold value of $0.03$ (top),
full multidimensional solution calculated on a two-dimensional grid (middle) and full VE model (bottom).
The simulation time is $t = 180$~h.}
\label{satRecons2}
\end{figure}

We compare the results from the adaptive multiphysics model with a threshold value of 0.03, full multidimensional model, and VE model in Figure \ref{satRecons2}.
The newly developed multiphysics model compares well with the full multidimensional model: the horizontal extent of the plume is represented correctly as is the diversion of the
gas around the low-permeability lense. Differences to the full multidimensional solution are evident in the VE subdomain region, where a higher brine phase saturation is
calculated in the plume. This could be improved by a pseudo-VE model that assumes a pseudo-residual brine phase saturation in the plume which
is reduced dynamically due to slow vertical drainage \cite{WRCR:WRCR23006}.
We note that the full multidimensional model does not necessarily give better solutions in the VE subdomain.
If the VE assumption is sufficiently valid, the VE model may be more accurate than the full multidimensional model since it does not rely
on a finite grid discretization in the vertical direction.
The full VE model leads to an underestimation of the horizontal extent of the plume because
it is assumed that the gas is in equilibrium with the brine phase at all locations. This leads to the gas phase entering the lense since the
entry pressure here is not different than in other parts of the domain. Because of the low permeability in the lense, large parts of the gas phase
that have entered the lense are retained there.\par
The adaptive multiphysics model is significantly faster than the full multidimensional model even though this
is a small test case. Table \ref{tab:numberOfCells} shows the average number of cells and the CPU times for the models.
\begin{table}
\caption{Relative average number of cells (compared to the full multidimensional model) and relative CPU times for
full VE model, adaptive multiphysics model
(with different threshold values $\epsilon$) and full multidimensional model.}
\centering
\begin{tabular}{|c | c c|}
\hline
 & relative average & relative \\
Model & number of cells & CPU time\\
 & [-] & [-]\\
\hline
Full VE & 0.008 & 0.003\\
Multiphysics $\epsilon_\mathrm{relPerm} = 0.06$ & 0.04 & 0.02\\
Multiphysics $\epsilon_\mathrm{relPerm} = 0.05$ & 0.11 & 0.05\\
Multiphysics $\epsilon_\mathrm{relPerm} = 0.04$ & 0.12 & 0.06\\
Multiphysics $\epsilon_\mathrm{relPerm} = 0.03$ & 0.19 & 0.12\\
Multiphysics $\epsilon_\mathrm{relPerm} = 0.02$ & 0.3 & 0.18\\
Multiphysics $\epsilon_\mathrm{relPerm} = 0.01$ & 0.41 & 0.22\\
Full multidimensional & 1 & 1\\
\hline
\end{tabular}
\label{tab:numberOfCells}
\end{table}
The speed-up in the adaptive model is attributed to the reduction in the number of computational cells.
It leads to a smaller linear system to be solved for the pressure step in the IMPES algorithm and thus lower computational costs.
Note that we expect the adaptive multiphysics model to be even more computationally efficient for larger, three-dimensional test cases,
because evaluating the coupling criterion and adapting the grid requires relatively little computational resources.\par
\begin{figure}[!htb]
\centering
\includegraphics[width=0.6\textwidth]{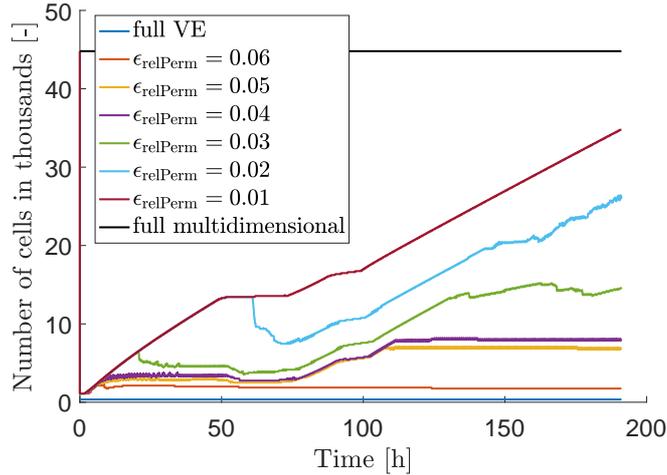}
\caption{Number of computational cells in the domain over simulated time
for full VE model, adaptive multiphysics model (with different threshold values $\epsilon$) and full multidimensional model.}
\label{numberOfCells}
\end{figure}

For different threshold values we plot the number of cells in the domain over simulated time in Figure \ref{numberOfCells}.
As expected, the number of computational cells is higher for lower threshold values at any time.
At the beginning of injection, the number of cells increases
as the plume advances regardless of which threshold value is chosen.
For higher threshold values, the number of cells remains stable as soon as a VE subdomain develops within the gas plume.
For the lower threshold value of 0.02 the plume develops a VE subdomain region only when the low-permeability lense is
reached and the gas phase is backed up. This is indicated by a drop in number of cells in the domain.
Until the end of the simulation the gas plume has not yet developed a VE subdomain after the lense, which is why the number of cells
is still increasing at the end of the simulation time for this threshold value and lower threshold values.
For an even lower threshold value of 0.01 the entire plume is discretized with the full multidimensional model over all times.
Since the plume advances into the domain over the entire time of simulation, the number of cells increases steadily for this threshold value.
Even so, a significant speed-up compared to a non-adaptive, full two-dimensional model is achieved,
since the one-phase region is a VE subdomain at all times. In many practical cases the extent of the plume in the horizontal plane
may be one or two orders of magnitude smaller than the domain and is locally restricted due to alternating injection and extraction cycles,
making the adaptive model even more favorable.\par
\subsection{Choice of threshold value for adaptive coupling}
\begin{figure}[!htb]
\centering
\includegraphics[width=0.7\textwidth]{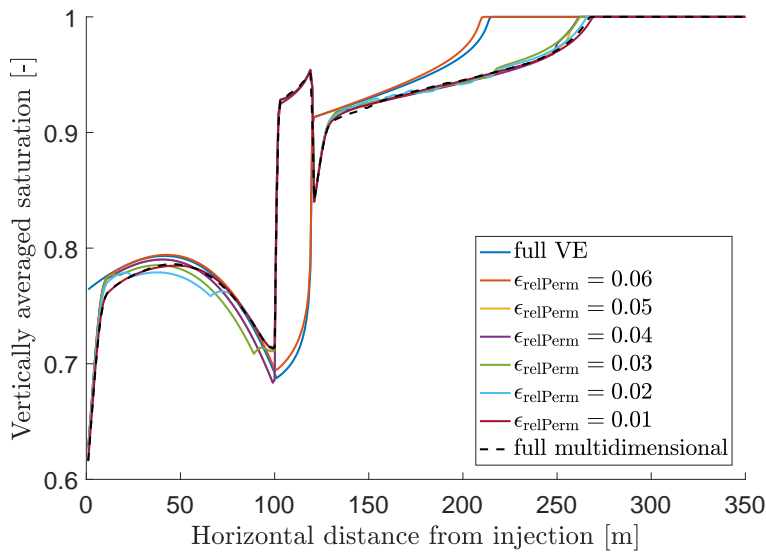}
\caption{Vertically averaged brine phase saturation over horizontal distance from injection location at $t = 192$~h
for full VE model, adaptive multiphysics model (with different threshold values $\epsilon$) and full multidimensional model.}
\label{satOverX}
\end{figure}

We analyze the influence of the threshold value and give recommendations for the choice of the threshold.
We plot vertically averaged brine phase saturation for the full VE model, for the multiphysics model with different threshold values
and the full multidimensional model in Figure \ref{satOverX}. For a very low threshold, where the entire plume is discretized with a full multidimensional
model, the results match very well with the full multidimensional model.
Differences with the full multidimensional model increase slightly
with an increase in threshold value,
especially for the averaged saturation in front of the low-permeability lense and the location of the subdomain boundaries.
At the subdomain boundaries the averaged brine phase saturation shows non-monotonic behavior with more gas in the VE subdomain than in the
full multidimensional subdomain. This is likely due to small differences between the two models at the subdomain boundary: the VE model
assumes vertical equilibrium of the two fluid phases which is not completely represented by the full multidimensional model at that location, either because of finite
grid size in the vertical direction or because the two phases are physically not in vertical equilibrium yet.
For a high threshold value, the low-permeability lense is not detected anymore and the results differ greatly from the previous results,
especially at the low-permeability lense and in the region behind it, as a consequence
of gas being trapped within the lense.
The same is observed for the full VE model, with an additional difference in averaged brine phase saturation close to the injection region.\par
\begin{figure}[!htb]
\centering
\includegraphics[width=0.65\textwidth]{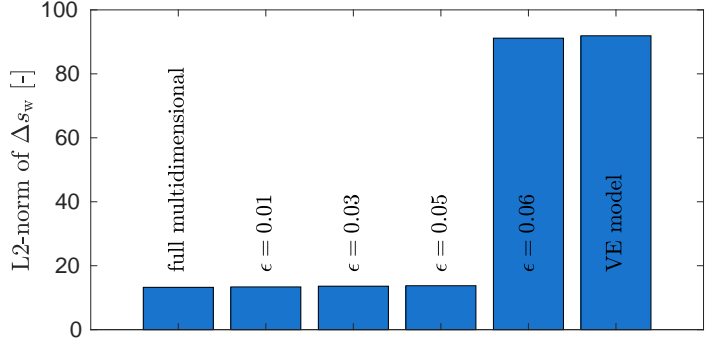}
\caption{L2-norm error of brine phase saturation at $t = 192$~h for full multidimensional model,
adaptive multiphysics model (with different threshold values $\epsilon$) and full VE model. Comparison is done
with respect to a full multidimensional reference on a very fine grid determined by grid convergence.}
\label{l2norm}
\end{figure}

We identify three major sources of errors in our models: applying the VE model in regions that are not in vertical equilibrium,
insufficient grid resolution especially in the vertical direction, and secondary effects due to the coupling.
The first error source is controlled by the threshold value while the second error source mainly applies to the full multidimensional model
and the full multidimensional subdomains in the multiphysics model.
In this context we analyze the influence of the threshold value on the accuracy of the adaptive multiphysics model.
We measure accuracy with the L2-norm error of the brine phase saturation.
The L2-norm error is calculated as the square root of the sum of squares of the saturation differences in each cell.
It is determined with respect to a full multidimensional reference that was obtained
on a very fine grid determined by grid convergence ($\Delta x = 0.25,\, \Delta y = 0.05$). The resolution of this grid is considered to be small enough so that
errors due to discretization are minimized but would be impracticable in real applications. It is used here to calculate the L2-norm error
for the full multidimensional model on the coarser and more practicable grid, the full VE model and the multiphysics model. This way, we can put
the accuracy of the multiphysics model into context of the discretization error
of the full multidimensional model, as shown in Figure \ref{l2norm} for a simulated time $t = 192$~h.
It can clearly be seen that the L2-norm error of the multiphysics model is similar to the L2-norm error of the full multidimensional model,
within a wide range of the threshold values.
Even when only small parts of the domain (injection area, advancing edge of the plume, low-permeability lense) are resolved with a
full multidimensional model, the multiphysics model still gives very accurate results.
This is because losses in accuracy of the multiphysics model due to the coupling that
leads, e.g., to the non-monotone vertically averaged saturation at the subdomain boundary observed in Figure \ref{satOverX},
are outbalanced by a better representation of the plume in the VE subdomains of the multiphysics model.
The coarse grid resolution in the vertical direction however leads to a significant L2-norm error for the full multidimensional model.
We can see that a large threshold value ($\epsilon=0.06$) leads to an L2-norm error close to that from the VE model,
which is significantly higher than for smaller threshold values ($\epsilon \leq 0.05$).
The jump in the L2-norm error indicates that in this case the multiphysics model fails to
accurately capture the relevant physical processes in the domain, like the gas flow around the low-permeability lense.
We note that this transition between L2-norms from one criterion value to another
is likely going to be smoother for a situation with more than one relevant local heterogeneity.
For a threshold value only slightly lower than this critical value (e.g., $\epsilon=0.05$),
the multiphysics model is much faster than the full multidimensional model
while showing the same accuracy, which make it a very efficient model.
\par
The choice of optimal threshold value depends on the specific problem,
the desired accuracy of simulation, and the full multidimensional grid
discretization.
An optimal threshold value can be determined by varying the threshold from larger to smaller values.
In analogy to a grid convergence study, we
determine the appropriate (average) size of the VE subdomain by adding step-wise more full multidimensional cells.
The threshold value has to be reduced in small steps so that a region for the threshold value can be identified
for which the results do not change significantly anymore with further reduction
of the threshold value.
This can be seen in Figure \ref{satOverX}, where for a very high threshold value the results are much different from all lower threshold values.
This approach may require multiple runs of the multiphysics model before the optimal threshold value is found.
However, starting with a high threshold for the multiphysics model results already in a very fast model and
once the optimal threshold value is found, a large number of most efficient simulation runs can be carried out, e.g., for a Monte-Carlo type simulation.
\section{Conclusions}
In this paper we have developed an adaptive multiphysics model that couples a full multidimensional model to a VE model.
The coupling is realized in a monolithic framework. We couple the fluxes over the subdomain boundaries
by using variables in the full multidimensional boundary cells and reconstructed fine-scale variables in the VE boundary subcells.
The unknown variables in the VE subcells are expressed as fine-scale reconstructions of the VE cell variables
using the assumption of vertical equilibrium.
The pressure and saturation equations are solved sequentially with an IMPES algorithm, where we solve the pressure
implicitly for the entire domain.
The subdomains are assigned adaptively during simulation based on a local, \textit{a posteriori} criterion for vertical equilibrium that compares
computed and reconstructed vertical profiles of saturation or relative permeability in the grid columns.\par
The adaptive multiphysics model showed high accuracy in predicting the gas distribution with a much smaller number of grid cells
and consequently lower computational cost compared to a full multidimensional model.
The multiphysics model can accurately capture full multidimensional flow dynamics, e.g., around the injection location or heterogeneities farther away,
and has a high accuracy in the VE subdomains where the VE assumption is valid.
The threshold value to determine the VE subdomains can be chosen by decreasing the threshold value step-wise in a test similar to grid convergence tests.
Overall, the multiphysics model coupling VE and full multidimensions is an efficient tool for modeling large scale applications of gas injections
in the underground.
We aim to extend the multiphysics model to include non-isothermal and compositional effects and will investigate
applications to more challenging problems of subsurface energy storage as part of the ongoing work.

\section*{Acknowledgments}
Beatrix Becker is supported by a scholarship of the Landesgraduiertenf\"orderung Baden-W\"urttemberg at the University of Stuttgart.
Beatrix Becker would like to gratefully acknowledge financial support of the research abroad granted by the German Research Foundation (DFG)
in the Cluster of Excellence in Simulation Technology (EXC 310/1) at the University of Stuttgart.

\pdfbookmark[0]{Bibliography}{Bibliography}
\bibliographystyle{plainurl}
\bibliography{literature}

\end{document}